\documentclass{appolb}
\usepackage{graphicx}

\begin{document}
\title{Impact of the underlying O(4) criticality on net-proton number fluctuations%
\thanks{Presented at the 45th Congress of Polish Physicists, Krakow, Poland, 13-18 September, 2019}%
}
\author{Micha\l{} Szyma\'nski
\address{Institute of Theoretical Physics, University of Wroclaw, Wroclaw, Poland}}
\maketitle

\begin{abstract}
We discuss the beam-energy dependence of
ratios of the first four cumulants of the net-proton number, calculated using a phenomenologically motivated model in which critical mode fluctuations couple to protons and anti-protons. The model takes into account an impact of the underlying $O(4)$ criticality on the  QCD phase diagram in the presence of a $Z(2)$ critical point. This allows us to capture qualitatively both the monotonic behavior of the
lowest-order ratio and the non-monotonic behavior of higher-order ratios which are seen in the experimental data from the STAR Collaboration. We also discuss the dependence of our results on the coupling strength between the critical mode and the protons as well as the location of the critical point.
\end{abstract}


\section{Introduction}

One of the unresolved questions of QCD phenomenology concerns the existence and location of the critical point (CP) in the $T$ and $\mu$ plane. Fluctuations of the critical mode ($\sigma$), although not directly measurable, are expected to affect fluctuations of conserved charges~\cite{Stephanov:1998dy,Stephanov:1999zu}. One of these is the baryon number, whose fluctuations, experimentally, are probed by net-proton number fluctuations. Preliminary results of the STAR Collaboration~\cite{Luo:2015ewa,Luo:2015doi,Thader:2016gpa} show a non-monotonic beam energy dependence of the ratios of higher order net-proton number cumulants. This data, however, is still not fully interpreted and therefore effective models are important tools to improve our understanding of these quantities.

One of such models was developed in~\cite{Bluhm:2016byc} and allowed for the qualitative description of the non-monotonic behavior of the $C_3/C_2$ and $C_4/C_2$ ratios. However, it also exhibited the strong non-monotonic behavior of the $C_2/C_1$ ratio which is not observed experimentally. Recently, that model was modified~\cite{Szymanski:2019yho} to take into account the impact of the underlying $O(4)$ criticality on the scaling properties of the baryon number and chiral susceptibilities~\cite{Hatta:2002sj,Sasaki:2006ws,Sasaki:2007qh} which allows for a better description of the $C_2/C_1$ ratio. We present ratios of net-proton number cumulants obtained using the refined model and discuss their dependence on the coupling strength between critical mode and (anti)protons and the location of the critical point.

\section{Model setup}

In this work, the regular contributions to net-proton number cumulants are calculated using the hadron resonance gas (HRG) model in which the QCD pressure is approximated by gas of non-interacting particles and their resonances. To include critical mode fluctuations in that model we follow the approach employed in Ref.~\cite{Bluhm:2016byc} and use the phenomenological relation for the particle mass, as suggested by linear sigma models,
\begin{equation}
m_i\sim m_0+g\sigma\,,   
\end{equation}
where $i=p,\bar{p}$ stands for (anti)proton, $m_0$ is a non-critical contribution to the mass and $g$ is the coupling strength between (anti)protons and the critical mode $\sigma$. Consequently, a distribution function becomes modified, such that
\begin{equation}
    f_i= f_i^0+f_i^1\delta \sigma\,,
\end{equation}
where $f_i^1$ can be found in Refs.~\cite{Bluhm:2016byc,Szymanski:2019yho}.

The n-th order cumulant of (anti)protons is given by
\begin{equation}
C^i_n=VT^3\frac{\partial^{n-1} (n_i/T^3)}{\partial(\mu_i/T)^{n-1} }\bigg\vert_{T=const.} \,,
\end{equation}
where $n_i$ is a number density and $\mu_i$ the corresponding chemical potential. In this work we consider the first four cumulants of the net-proton number, $N_{p-\bar{p}}=N_p-N_{\bar{p}}\,$, which, after neglecting the contribution of resonances and their decays, read~\cite{Bluhm:2016byc}
\begin{eqnarray}\label{eq:c_n_original}
C_n&=&C_n^p+(-1)^nC_n^{\bar{p}}+(-1)^n\langle(V\delta\sigma)^n\rangle_c (m_p)^n (J_p-J_{\bar{p}})^n\,,
\end{eqnarray}
where $C_n^p$ and $C_n^{\bar{p}}$ are n-th order proton and anti-proton cumulants obtained within HRG model, respectively, $\langle(V\delta\sigma)^n\rangle_c$ is n-th cumulant of the critical mode and $J_i$ is the $\sigma$-independent factor.

Under the assumption that QCD and three-dimensional Ising model belong to the same universality class, the QCD order parameter close to the CP can be identified with the magnetization, $M_I$, and the critical mode cumulants can be written as~\cite{Bluhm:2016byc}
\begin{equation}
\langle(V\delta\sigma)^n\rangle_c=\left(\frac{T}{VH_0}\right)^{n-1}\left.\frac{\partial^{n-1}M_I}{\partial h^{n-1}}\right\vert_r\,,
\label{eq:m_cumulant}
\end{equation}
where $r$ and $h$ are the spin model reduced temperature and magnetic field, which are mapped on the QCD temperature and baryon chemical potential. A detailed discussion of the mapping as well as the magnetic equation of state is presented in the papers~\cite{Bluhm:2016byc,Szymanski:2019yho}.

\begin{figure}
\centerline{
\includegraphics[width=12.5cm]{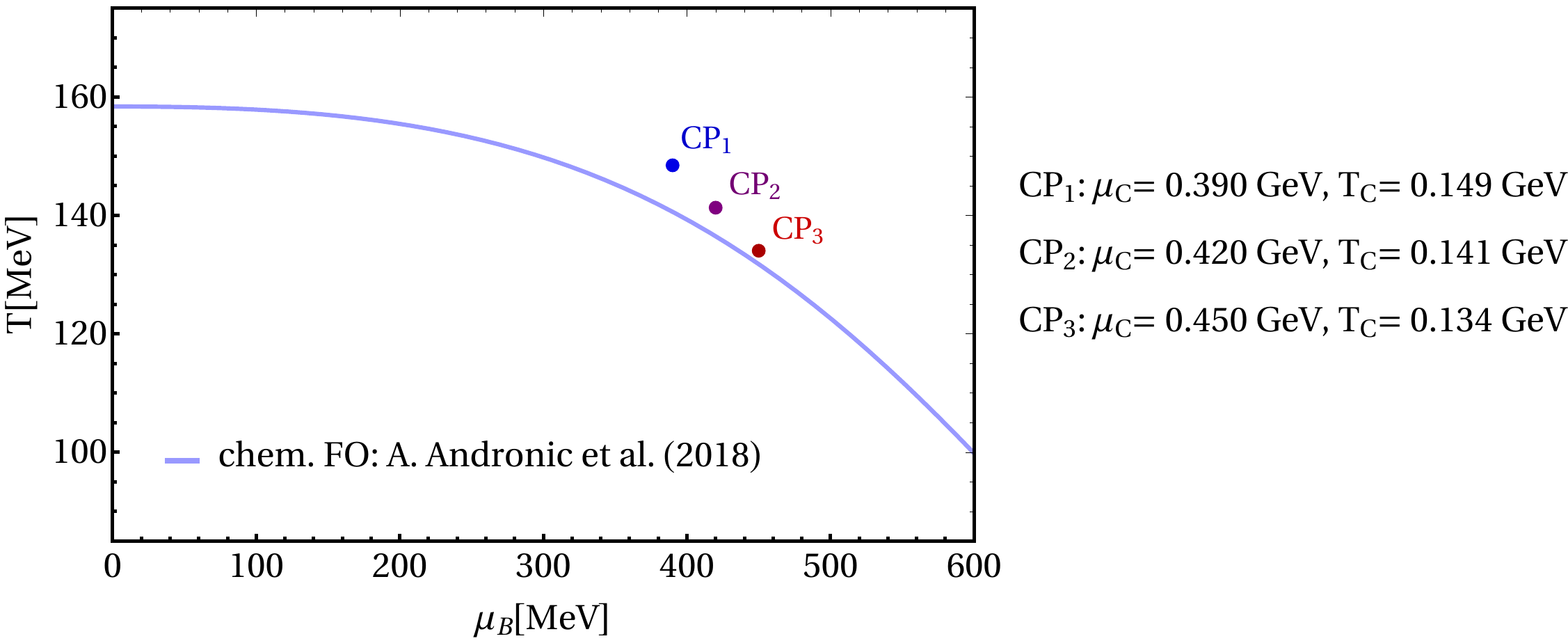}}
\caption{(Color online) The locations of QCD critical point considered in the current study, together with the recently obtained chemical freeze-out curve~\cite{Andronic:2017pug}. The lines of first order phase transitions are not shown.}
\label{fig:cp_setup}
\end{figure}

In this approach we find that (see Eq. \ref{eq:c_n_original}),
\begin{equation}
C_2^{\rm{sing.}} \propto \frac{\partial M_I}{\partial h}\sim\chi_{\mathrm{chiral}},
\end{equation}
where $\chi_{\mathrm{chiral}}$ is the chiral susceptibility of QCD and the second relation follows from the universality. Such a relation, expected in the $Z(2)$ universality class, holds only very close to CP~\cite{Hatta:2002sj}. Further away the traces of the tricritical point make the chiral susceptibility diverge stronger than the baryon number one~\cite{Hatta:2002sj,Sasaki:2006ws,Sasaki:2007qh}. To take that effect into account, the model introduced in Ref.~\cite{Bluhm:2016byc} was modified~\cite{Szymanski:2019yho}. The modification is based on the following relation obtained within the effective model calculations for $O(4)$ and tricritical points~\cite{Sasaki:2006ws,Sasaki:2007qh},
\begin{equation}
\chi_{\mu\mu}\simeq\chi_{\mu\mu}^{reg}+\mathbf{\sigma}^2\chi_{\rm{chiral}}\,,
\end{equation}
where $\chi_{\mu\mu}^{reg}$ is the regular part of the baryon number susceptibility. Such a form of the second cumulant can be obtained by replacing the proton mass in Eq. \ref{eq:c_n_original} with the order parameter, $g\sigma$. After the modification, we find that
\begin{equation}
C_2=C_2^p+C_2^{\bar{p}}+g^2\sigma^2 \langle(V\delta\sigma)^n\rangle (J_p-J_{\bar{p}})^2\,.
\label{eq:c2_model}
\end{equation}
We modify higher order cumulants accordingly:
\begin{eqnarray}
C_3&=&C_3^p-C_3^{\bar{p}}- g^3\sigma^3 \langle(V\delta\sigma)^n\rangle (J_p-J_{\bar{p}})^3\,,\label{eq:c3_model}\\
C_4&=&C_4^p+C_4^{\bar{p}}+g^4\sigma^4 \langle(V\delta\sigma)^n\rangle (J_p-J_{\bar{p}})^4\,.
\label{eq:c4_model}
\end{eqnarray}
These quantities are volume-dependent, and hence it is convenient to consider their ratios, in which this dependence cancels out,
\begin{equation}
\frac{C_2}{C_1}=\frac{\sigma^2}{M}\,,\qquad \frac{C_3}{C_2}=S\sigma\,,\qquad \frac{C_4}{C_2}=\kappa\sigma^2\,,
\label{eq:ratios}
\end{equation}
where $M$ is the mean, $\sigma^2$ the variance, $\kappa$ the kurtosis and $S$ the skewness.  To compare model results with the experimental data on event-by-event multiplicity fluctuations we calculate the net-proton number cumulants at the chemical freeze-out using the recently obtained parametrization~\cite{Andronic:2017pug} (see the blue line on Fig. \ref{fig:cp_setup}).

\begin{figure}
\includegraphics[width=6.25cm]{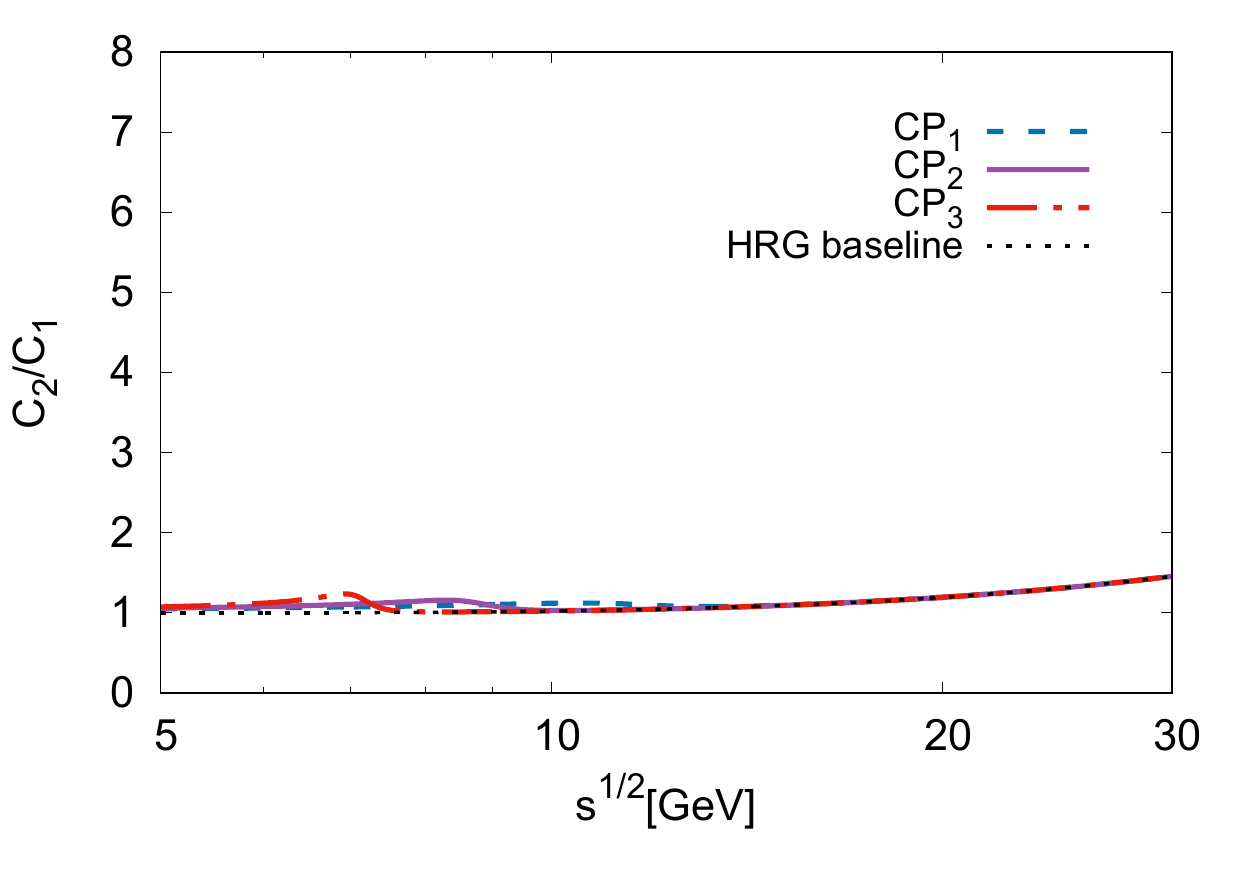} \includegraphics[width=6.25cm]{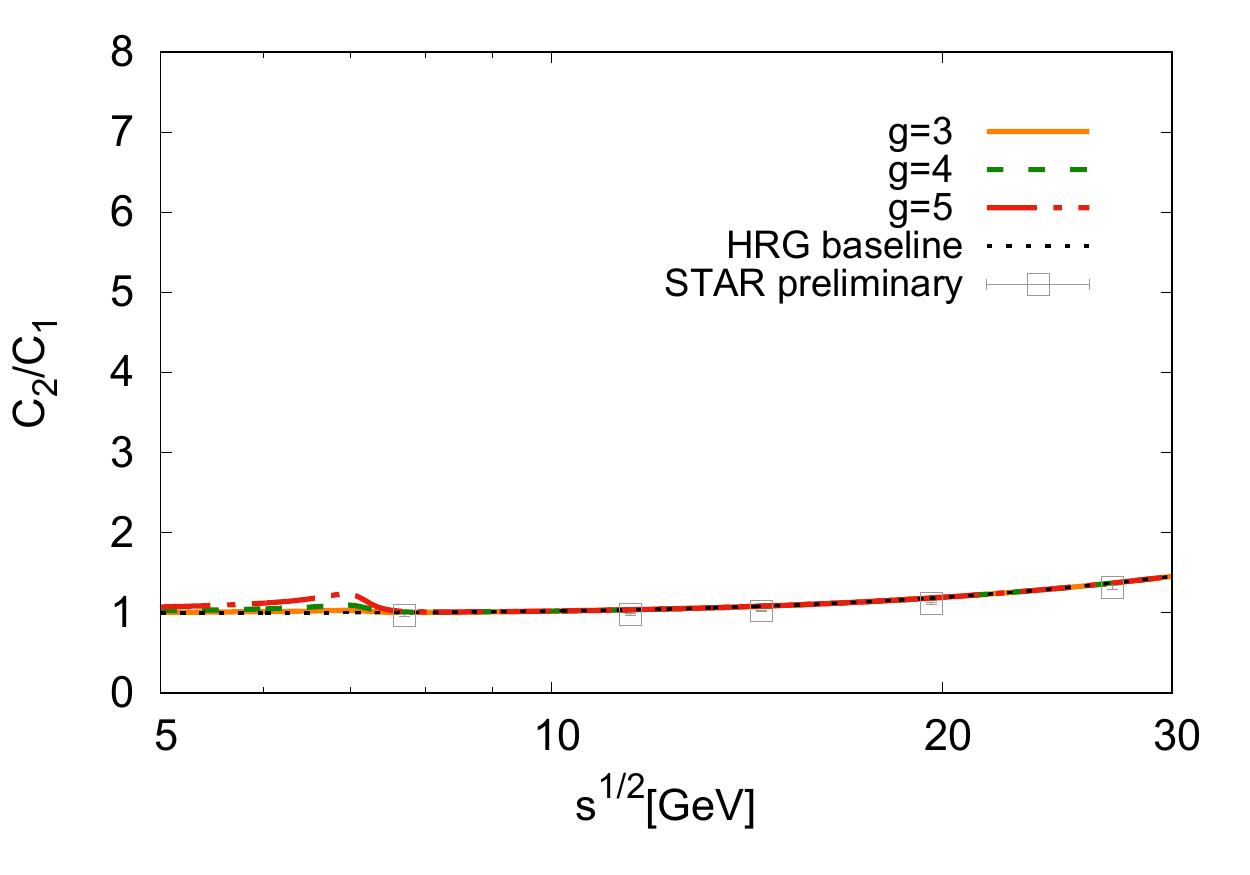}\\
\includegraphics[width=6.25cm]{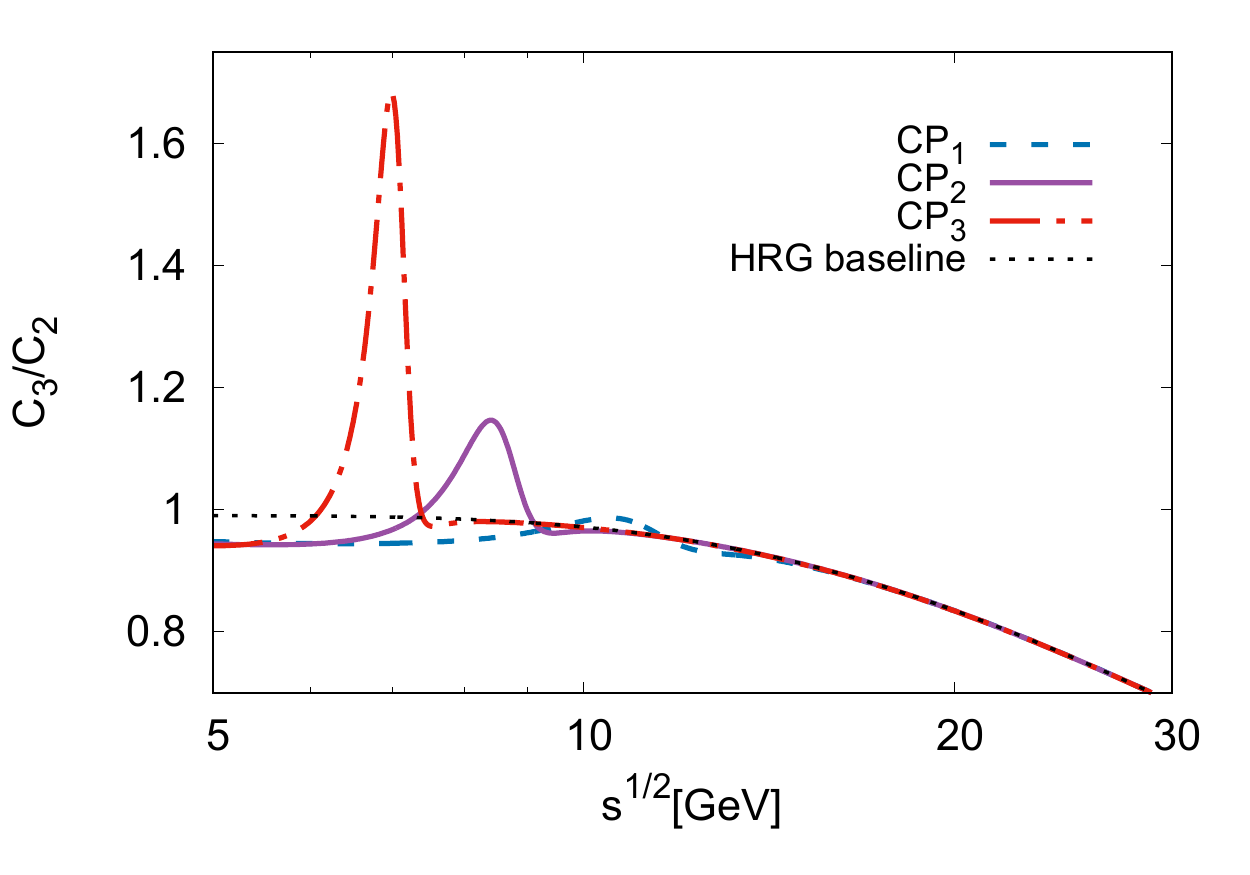} \includegraphics[width=6.25cm]{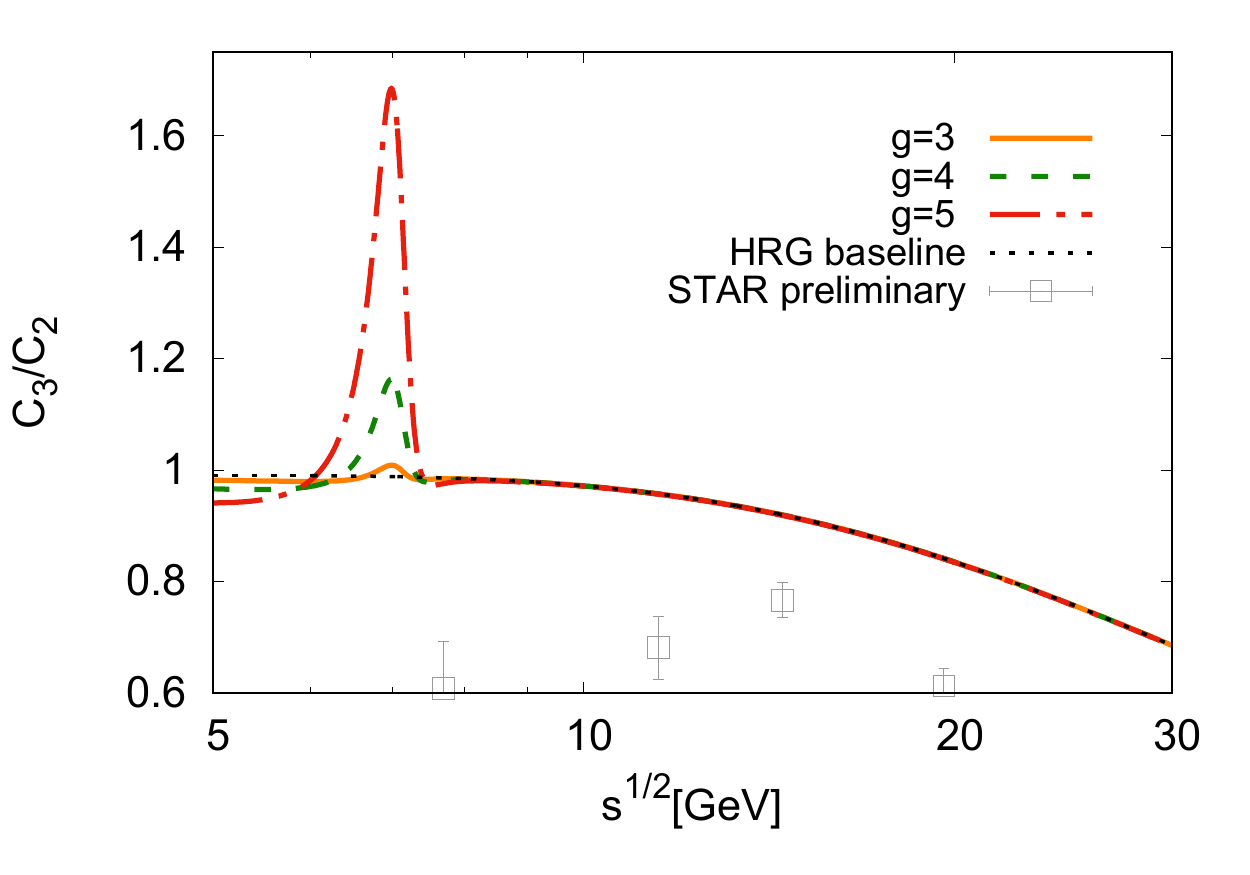}\\
\includegraphics[width=6.25cm]{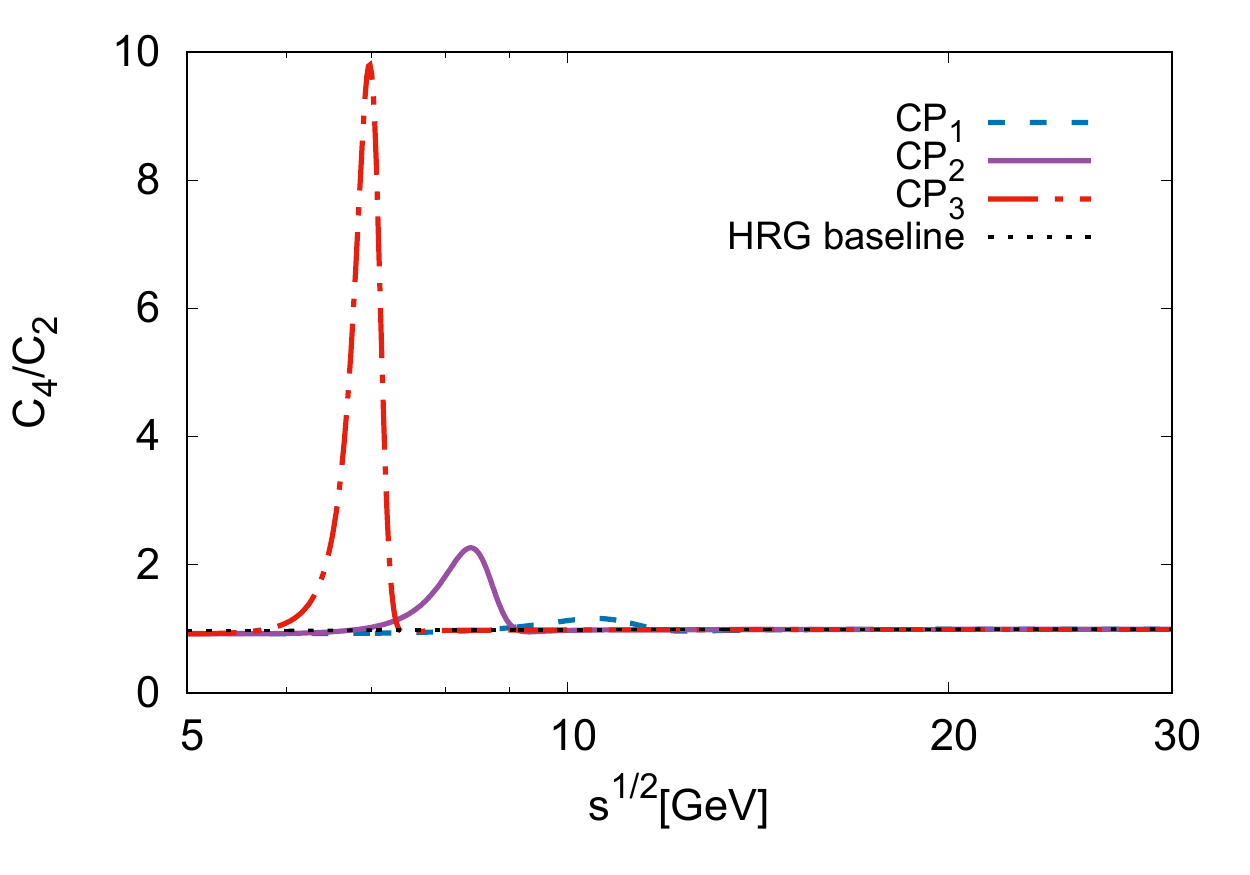} \includegraphics[width=6.25cm]{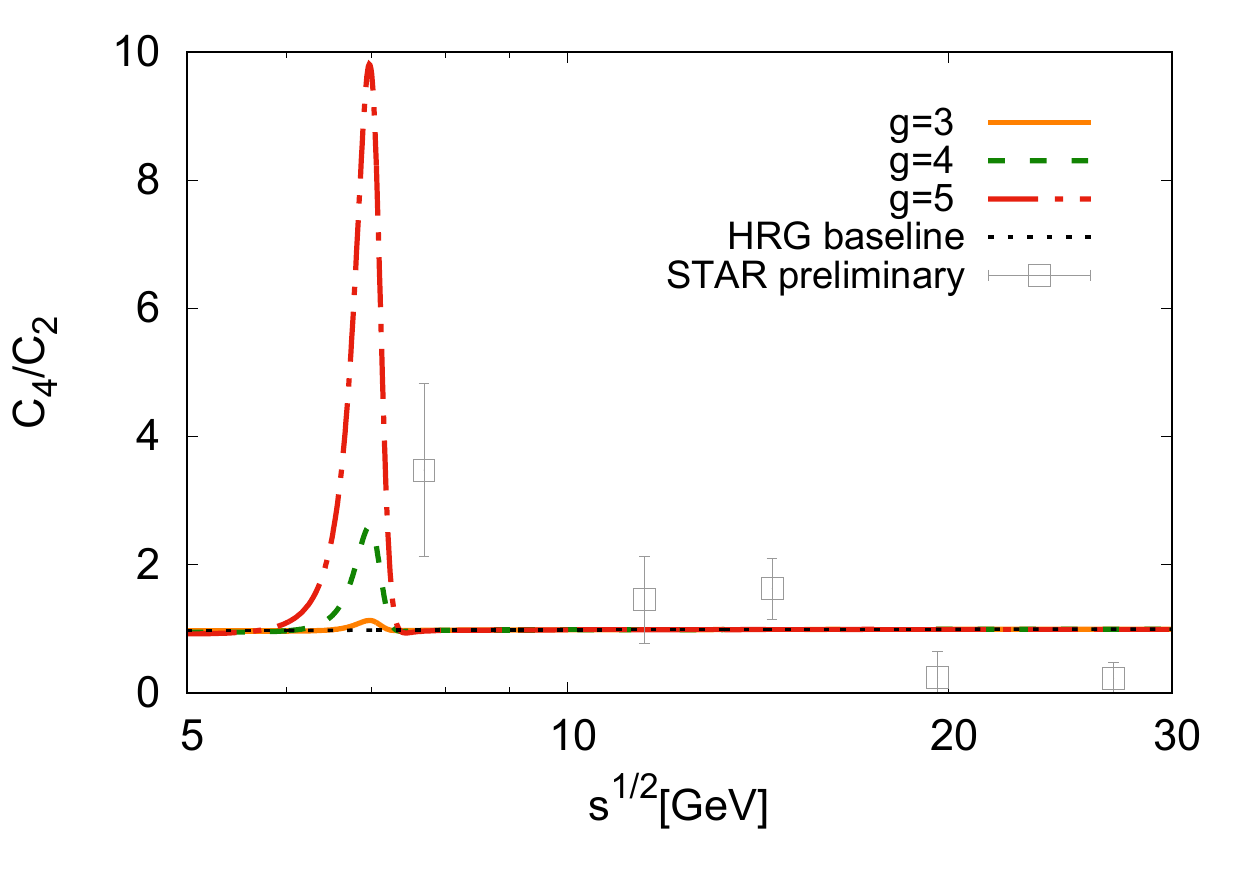}
\caption{(Color online) Ratios of net-proton number cumulants calculated for the fixed coupling $g=5$ and for different locations of the QCD critical point, as shown on Fig.~\ref{fig:cp_setup}, (the left column) and for the fixed critical point location $CP_3$ for different values of the coupling strength $g=2,$ 3 and 5 (the right column).
}
\label{fig:results}
\end{figure}

\section{Numerical results}

To study the effect of the position of the critical point in the QCD phase diagram on cumulant ratios we consider three different locations of CP which are shown in Fig. \ref{fig:cp_setup}, where the distance from the freeze-out curve is largest for CP$_1$ and smallest for $CP_3$. A detailed discussion of the remaining model parameters and their values can be found in Refs.~\cite{Bluhm:2016byc,Szymanski:2019yho}.

The left column of Fig. \ref{fig:results} shows the net-proton number cumulant ratios obtained for different locations of the critical point (as shown in Fig. \ref{fig:cp_setup}) with a fixed value of coupling, $g=5$. A non-monotonic behavior of cumulant ratios becomes more pronounced when the critical point is closer to the freeze-out line and the deviation from the non-critical HRG baseline becomes larger for higher order cumulant ratios. The right column of Fig.~\ref{fig:results} shows the coupling strength dependence of net-proton number cumulant ratios for CP$_3$. When compared to the STAR data~\cite{Thader:2016gpa}, we find a qualitative agreement between our model and experimental results for the $C_2/C_1$ and $C_4/C_2$ ratios. On the other hand, the model result on the $C_3/C_2$ ratio does not follow the systematics seen in the experimental data. The former are above the HRG baseline while the latter stay below.

Therefore, with the proper choice of parameters, our model allows to describe some of the experimentally observed cumulant ratios. Especially, the smooth dependence of $C_2/C_1$ as well as strong increase of $C_4/C_2$ at low beam energies, $\sqrt{s}<20\,$GeV, suggest that the QCD critical point may be located close to the phenomenological freeze-out curve. In this case, the $C_3/C_2$ ratio should be also above the non-critical baseline. However, this is not seen in the experimental data and  therefore it seems unlikely that the QCD critical point is close to the freeze-out curve.

\section{Conclusions}

We studied ratios of net-proton number cumulants obtained within an effective model which takes into account the effect of the underlying $O(4)$ criticality. Model results were compared with the recent experimental data on net-proton number fluctuations from the STAR Collaboration. Our model allows to describe some of the experimentally observed features in the net-proton number cumulant ratios. Particularly, smooth dependence of $C_2/C_1$ and rise of $C_4/C_2$ at low beam energies suggest that the critical point may be located close to the freeze-out curve. However, the experimentally observed $C_3/C_2$ ratio does not follow the behavior expected from such a scenario. 

Therefore, it seems unlikely that the critical point is located close to the phenomenological freeze-out curve.  However, this statement requires further investigation because of current uncertainties on both theoretical as well as experimental sides.\\[0.4cm]
The author acknowledges the stimulating discussions and fruitful collaboration with M. Bluhm, K. Redlich and C. Sasaki. This work is partly supported by the Polish National Science Center (NCN) under the Maestro grant DEC-2013/10/A/ST2/00106, Opus grant 2018/31/B/ST2/01663 and Polonez grant UMO-2016/21/P/ST2/04035. We thank F.~Geurts and J.~Th\"ader for providing the preliminary STAR data~\cite{Thader:2016gpa} shown in Figure~\ref{fig:results}.


\begin{thebibliography}{99}

\bibitem{Stephanov:1998dy}
  M.~A.~Stephanov, K.~Rajagopal and E.~V.~Shuryak,
  Phys.\ Rev.\ Lett.\  {\bf 81}, 4816 (1998).
  
\bibitem{Stephanov:1999zu}
  M.~A.~Stephanov, K.~Rajagopal and E.~V.~Shuryak,
  Phys.\ Rev.\ D {\bf 60},  114028 (1999).

\bibitem{Luo:2015ewa}
  X.~Luo [STAR Collaboration],
  PoS CPOD {\bf 2014}, 019 (2015).

\bibitem{Luo:2015doi}
  X.~Luo,
  Nucl.\ Phys.\ A {\bf 956}, 75 (2016).

\bibitem{Thader:2016gpa}
  J.~Th\"ader [STAR Collaboration],
  Nucl.\ Phys.\ A {\bf 956} 320 (2016).
 
\bibitem{Bluhm:2016byc}
  M.~Bluhm, M.~Nahrgang, S.~A.~Bass and T.~Sch\"afer,
  Eur.\ Phys.\ J.\ C {\bf 77}, no. 4, 210 (2017).
  
\bibitem{Szymanski:2019yho} 
  M.~Szyma\'{n}ski, M.~Bluhm, K.~Redlich and C.~Sasaki,
  \textit{arXiv}:1905.00667 [nucl-th], to appear in J. Phys. G

\bibitem{Hatta:2002sj}
  Y.~Hatta and T.~Ikeda,
  Phys.\ Rev.\ D {\bf 67}, 014028 (2003).

\bibitem{Sasaki:2006ws}
  C.~Sasaki, B.~Friman and K.~Redlich,
  Phys.\ Rev.\ D {\bf 75}, 054026 (2007).

\bibitem{Sasaki:2007qh}
  C.~Sasaki, B.~Friman and K.~Redlich,
  Phys.\ Rev.\ D {\bf 77}, 034024 (2008).


\bibitem{Andronic:2017pug} 
  A.~Andronic, P.~Braun-Munzinger, K.~Redlich and J.~Stachel,
  Nature {\bf 561}, no. 7723, 321 (2018)
  

\end{thebibliography}
\end{document}